\documentclass[10pt,journal,compsoc]{IEEEtran} %twoside

\usepackage[hidelinks]{hyperref}
\usepackage{amsmath,amssymb,amsfonts}
\usepackage{algorithmic}
\usepackage{graphicx}
\usepackage{textcomp}
\usepackage{array,multirow}
\usepackage[linesnumbered,lined,commentsnumbered]{algorithm2e}%
\usepackage{xcolor}
\usepackage[backend=bibtex,style=ieee,sorting=none,dashed=false]{biblatex}
\usepackage[inline]{enumitem}
\usepackage{anyfontsize}
\usepackage{makecell}
\usepackage{array}
\usepackage{soul}
\usepackage[utf8]{inputenc}

\usepackage{fancyhdr}

\usepackage{CJKutf8}

\definecolor{gr}{gray}{0.35}

\bibliography{bibliography}

\SetNlSty{bfseries}{\color{black}}{}

\ifCLASSINFOpdf
\else
\fi

\begin{document}
\title{FLiMS: a Fast Lightweight 2-way\\Merger for Sorting} %
\author{Philippos~Papaphilippou, Wayne~Luk and Chris~Brooks
\IEEEcompsocitemizethanks{\IEEEcompsocthanksitem The authors Philippos Papaphilippou and Wayne Luk are with the Department of Computing, Imperial College London, UK (E-mail: p.p.cy@ieee.org, w.luk@imperial.ac.uk)}
\IEEEcompsocitemizethanks{\IEEEcompsocthanksitem Chris Brooks is with Science Innovation, dunnhumby, UK (E-mail: Chris.Brooks@dunnhumby.com)}
}

\IEEEtitleabstractindextext{%
\begin{abstract}

In this paper, we present FLiMS, a highly-efficient and simple parallel algorithm for merging two sorted lists residing in banked and/or %
wide
memory. On FPGAs, its implementation uses fewer 
hardware resources than the state-of-the-art alternatives, due to the reduced number of comparators and elimination of redundant logic found on prior attempts. 
In combination with the distributed nature of the selector stage, a higher performance is achieved for the same amount of parallelism or higher. This is useful in many applications such as in parallel merge trees to achieve high-throughput sorting, where the resource utilisation of the merger is critical for building large trees and internalising the workload for fast computation. Also presented are efficient variations of FLiMS for optimizing throughput for skewed datasets, achieving stable sorting or using fewer dequeue signals. Additionally, FLiMS is shown to perform well as conventional software on modern CPUs supporting single-instruction multiple-data (SIMD) instructions, surpassing the performance of some standard libraries for sorting.

\end{abstract}

\begin{IEEEkeywords}
FPGA, high-throughput, parallel merger, merge tree, sorting algorithms,
sorting networks, SIMD, databases.
\end{IEEEkeywords}

}
\maketitle

\IEEEpeerreviewmaketitle

\section{Introduction}

\IEEEPARstart{T}{he} 
merge operation is widely used for a variety of applications, including in many popular sorting algorithms, such as mergesort and timsort \cite{auger2015merge}, as well as in hardware for database operations, including sort-merge joins \cite{mergejoin}. It is thus desirable to accelerate, and recent research has proposed a variety of merge accelerators on field-programmable gate arrays (FPGAs).

As frequency scaling has stopped being the primary method for achieving performance, the main way of achieving high-throughput/ high-bandwidth in modern systems is now to increase the datapath width. This has influenced computer architecture in many aspects, such as with wider single-instruction multiple-data (SIMD) instructions on general purpose processors (CPUs). One application that can benefit from high-throughput is sorting, as more data are being able to be processed per cycle. This paper presents a high-throughput merger algorithm, that merges two arbitrarily long input lists with high-throughput, exporting \(w\) elements per cycle, assuming the inputs are appropriately provided, such as through banked memories, reading up to \(w\) elements from each of the two input lists per cycle.

The 
challenges for the merge operations on FPGAs have been the low clock frequency due to the feedback datapath %of the merger consisting 
being the critical path, and the high resource utilisation in some attempts to remove the feedback datapath. %

At the time of writing, FLiMS uses the least number of comparators and pipeline stages among the latest mergers. It uses a modified version of the bitonic merge block, as found in bitonic sorters (see figure \ref{bits}), repurposed for performing 2-way parallel merge for streaming data. All alternative designs require the two input sequences of the bitonic (partial) merger (or the odd-even merge-based equivalent) to be sorted. The \emph{main idea} is to relax this condition. This idea eliminates the need for expensive rotations before the inputs of the merger \cite{pmt}, or alternative workarounds involving redundant logic \cite{mms,vms,ehms}. 

Other novel aspects in this work include the distributed nature of the selector stage, which has better timing characteristics on FPGAs, without occupying additional pipeline stages from the merger logic. The optimisation for skewed datasets is more lightweight and scalable than a previous attempt \cite{pmt}, as it does not rely on barrel shifters. FLiMS also does not suffer from the tie-record challenge found in all other feedback-less designs \cite{mms,vms,ehms, ehmsp}, and a costly workaround is not deemed necessary.  The regularity in the topology of the comparators %
is also found to help with SIMD implementations of high-throughput merging in CPUs.

The main contribution is the highly-efficient design of a high-throughput 2-way merger. Three variations of FLiMS are presented to achieve stable merging, high-throughput on skewed datasets and efficient memory use for special cases. Other contributions include a systematic comparison with a variety of alternative approaches, and an evaluation on both FPGAs and modern CPUs with SIMD. Automated generator scripts\footnote{{\emph{\textbf{Source available:} \url{http://philippos.info/flimsj}}}} provide (a) the Verilog code for the FPGA implementation, as well as (b) C++ code with SIMD intrinsics for CPUs, for a user-specified degree of parallelism (\(w\)). 

This paper is an extension of ``FLiMS: Fast Lightweight Merge Sorter'' \cite{flims} and is also partly based on the paper ``An Adaptable High-Throughput FPGA Merge Sorter for Accelerating Database Analytics'' \cite{fsorter}. %
It extends the material from these two papers with more detailed proofs (section \ref{corr}) and a thorough comparison with related work, such as providing exact values for the number of comparators each approach requires (section \ref{flimscomp}). Additional variations are presented to also support stable sorting and dequeuing whole rows (section \ref{afunc}). The experimental comparison is updated with implementations of the state-of-the-art alternatives \cite{ehms,ehmsp} on a high-end FPGA (section \ref{fpgaimplem}), and the SIMD-based version of FLiMS is extended to implement full sorting on modern CPUs to compare with sorting library functions (section \ref{avxflims}).

\section{Background}

On FPGAs, the available sorting accelerators are inspired by a variety of different serial algorithms, including insertion sort \cite{fpgasort,linear,fsorter}. However, this background section is restricted to high-throughput merge sorting on FPGAs. Merge sorting has been one of the most studied algorithms for sorting on FPGAs. This is due to its versatility, such as when reusing the same circuits to sort arbitrarily long input recursively. %

\subsection{High-throughput merge sorters}\label{bhtms}

High-throughput merge sorters can merge a number of sorted lists simultaneously, while providing an output rate of more than 1 element per cycle. This can be achieved by building a merge tree (PMT \cite{pmt}), mainly consisting of high-throughput mergers of 2 lists %
and FIFO queues, as with predecessors \cite{casper,kobayashi2015face}.  

\begin{figure}[h!]
\centering
\includegraphics[width=0.52\textwidth, trim=5 30 -5 10]{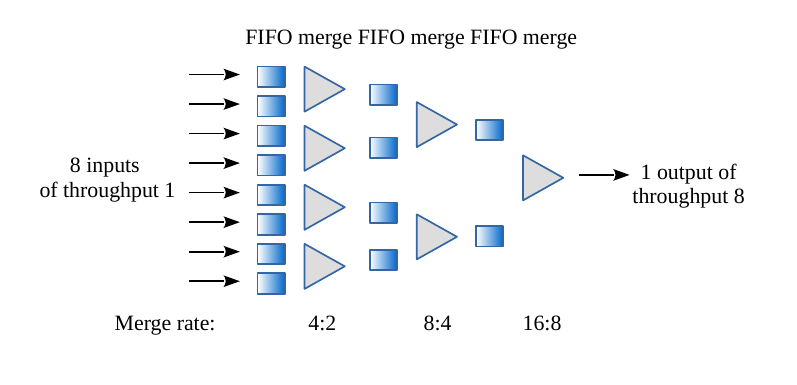}
\caption{Parallel merge tree (PMT \cite{pmt}), for 8 input lists}\label{pmtsf}
\end{figure}

Figure \ref{pmtsf} shows how these building blocks can be arranged to merge 8 sorted inputs of throughput 1, with an output rate 8. The `merge rate' of the mergers in each level of the merge tree denotes the number of elements in their two inputs collectively and the number of elements in their output. This directly contributes to the throughput of the merger, as well as the bandwidth requirements of the proceeding and succeeding logic in the pipeline. For instance, a merger of rate 4:2 merges two inputs of width 2 (times the element width) and outputs two elements per cycle. The difference in widths from level to level is managed by rate converters and the appropriate stall signals.

One challenge in parallel merge trees is that, while they can easily saturate the available bandwidth by scaling the number of inputs, they do not scale well for high number of inputs. For this reason, many-leaf mergers have emerged, to support a higher number of inputs more efficiently (currently up to a few thousands \cite{leaf,christopher} rather than in tens for PMTs). Merging many inputs simultaneously helps reduce the number of data passes required for complete sorting. A single data pass is equivalent to reading the entire input data once. However, many-leaf mergers are single-rate, meaning that they can only produce one output per cycle. If the data are not wide enough, this can lead to underutilisation of the available bandwidth. 

\begin{figure}[h!]
\centering
\includegraphics[width=0.52\textwidth, trim=5 30 -5 10]{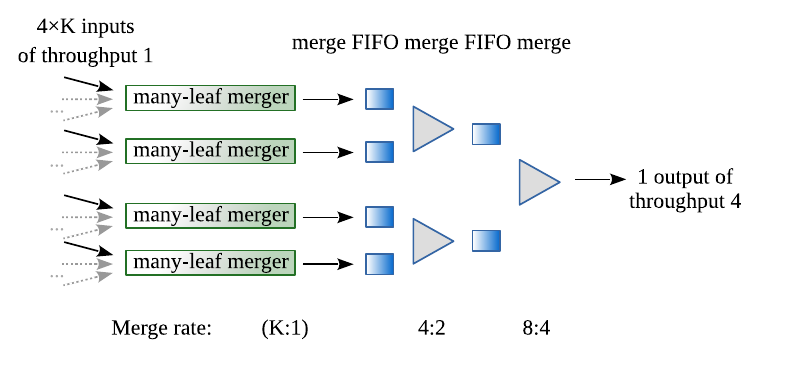}
\caption{HPMT for high-throughput and many-leaf merging}\label{hpmtsf}
\end{figure}

To solve these trade-offs, the hybrid parallel merge tree (HPMT) was introduced \cite{fsorter}, to enable both high-throughput and many-leaf merging at the same time. The size of the HPMT can be easily adjusted to saturate the bandwidth of the target architecture, while eliminating the number of passes of the data by still using many-leaf merging. Figure \ref{hpmtsf} shows how HPMT can combine 4 many-leaf mergers of \emph{K} inputs (totalling 4\(K\) input lists) into a merge tree with an output rate of 4 elements per cycle. %

\subsection{High-throughput 2-way mergers}\label{ht2ms}

The high-throughput 2-way mergers are the main building blocks of the aforementioned merge trees.

A merger for 2 already-sorted sublists of fixed length can be modified to merge 2 lists of arbitrary length in streaming fashion. Then it can be used as a building block for a parallel merge tree to merge many lists concurrently. %simultaneously.

Most of them are based on the two popular sorting networks: Batcher's odd-even mergesort and the bitonic sorter \cite{batcher}. %
These two sorting networks have the same number of stages and can be built hierarchically using 2 sorters of half the input and an appropriately sized merger to merge two equally-sized sorted sublists. The merger part consumes the last \(log_2(n)\) stages in both sorting networks, where \(n\) is the number of inputs. 
This merge block is optimised and/or combined with additional logic to work as a parallel merger for longer lists as streams. 

\begin{figure}[h!]
\centering
\includegraphics[width=0.46\textwidth, trim=0 10 -5 10]{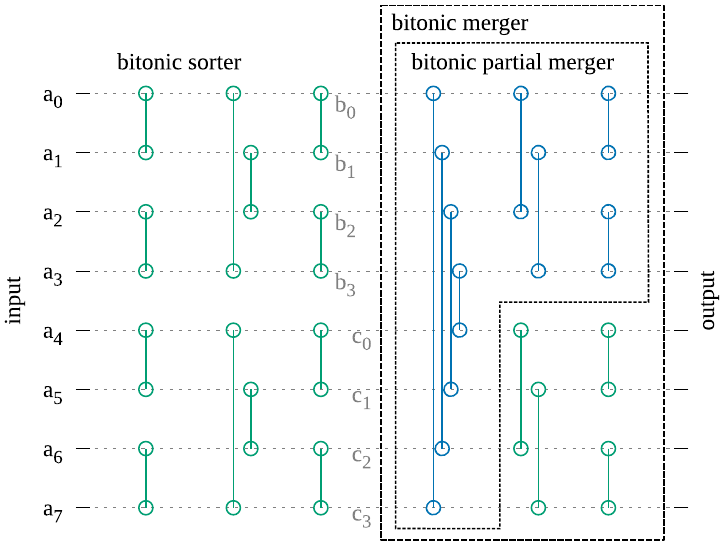}
\caption{A 8-to-4 bitonic partial merger as found in the bitonic sorter}
\label{bits}
\end{figure}

Figure \ref{bits} shows the bitonic merger, as found in a bitonic sorter of 8 inputs. The pairs of circles are compare-and-swap (\emph{CAS}) units, working as sorters of two inputs, i.e. \(a, b \rightarrow min(a, b), max(a, b)\). The list \([a_0, a_1, ..., a_7]\) is initially unsorted, and right before the merger, it is partially sorted into two sorted sublists (\([b_0, b_1, b_2, b_3]\) and \([c_0, c_1, c_2, c_3]\)). The merger merges these two sublists, consuming the last 3 stages of this bitonic sorter.

\begin{figure}[h!]
\centering
\includegraphics[width=0.3\textwidth, trim=0 10 0 10]{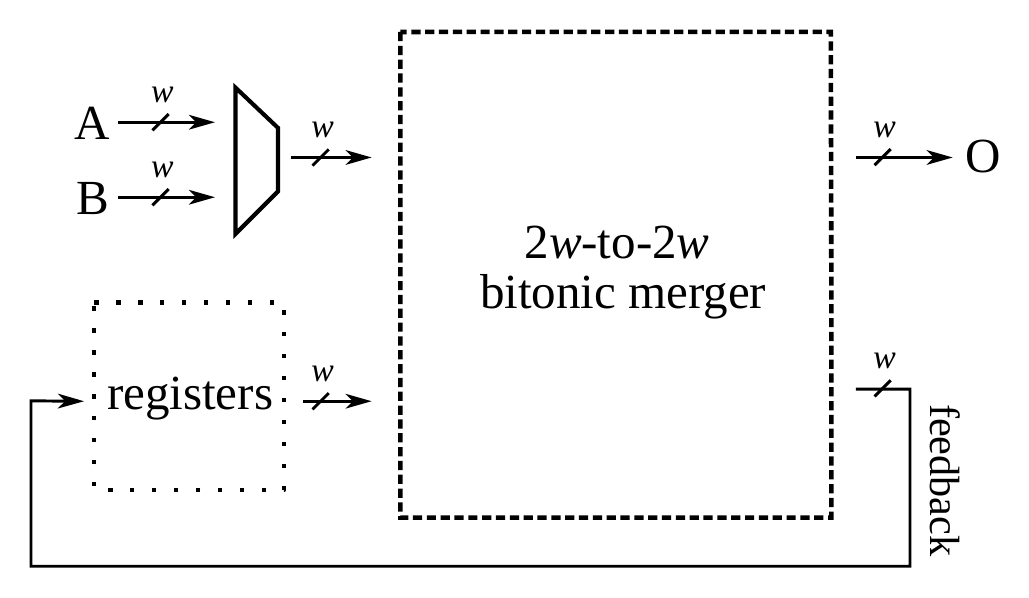}\\
\caption{Merger used in Casper et al. \cite{casper, simd2008}}\label{mergefirstl}
\end{figure}

The first known high-throughput merger for arbitrarily long input was based on a well known algorithm for merging using SIMD instructions on Intel processors \cite{simd2008}. It was adopted for database use on FPGAs \cite{casper}. However, the long feedback problem was more prominent on FPGAs \cite{mms}, since it can negatively impact the critical path, and is not scalable for many inputs. The algorithm goes as follows: starting with the first \(w\)-sized batches of each of the sorted sublists, the merger produces the top \(w\) result as output to progressively merge the entire input, while the lower \(w\) result is fed back into the lower \(w\) of the input to continue the merging. A single comparison between the first element of each batch is enough to distinguish the next batch to dequeue and place at the merger. %
Figure \ref{mergefirstl} shows the high-level representation of this approach for FPGAs.

Some 2-way hardware mergers (including FLiMS) utilise an optimization of the bitonic merger, the 2\(w\)-to-\(w\) bitonic partial merger \cite{farmahini2012modular}, which outputs just the top half of the result (or lower if the \emph{CAS} units are flipped). The bitonic partial merger is a subset of the bitonic merger. These are summarised in figure \ref{bits}. This ``pruned'' merge block is combined with additional logic to work as a parallel merger for longer lists as streams.

\begin{figure}[h!]
\centering
\includegraphics[width=0.34\textwidth, trim=0 20 0 20 ]{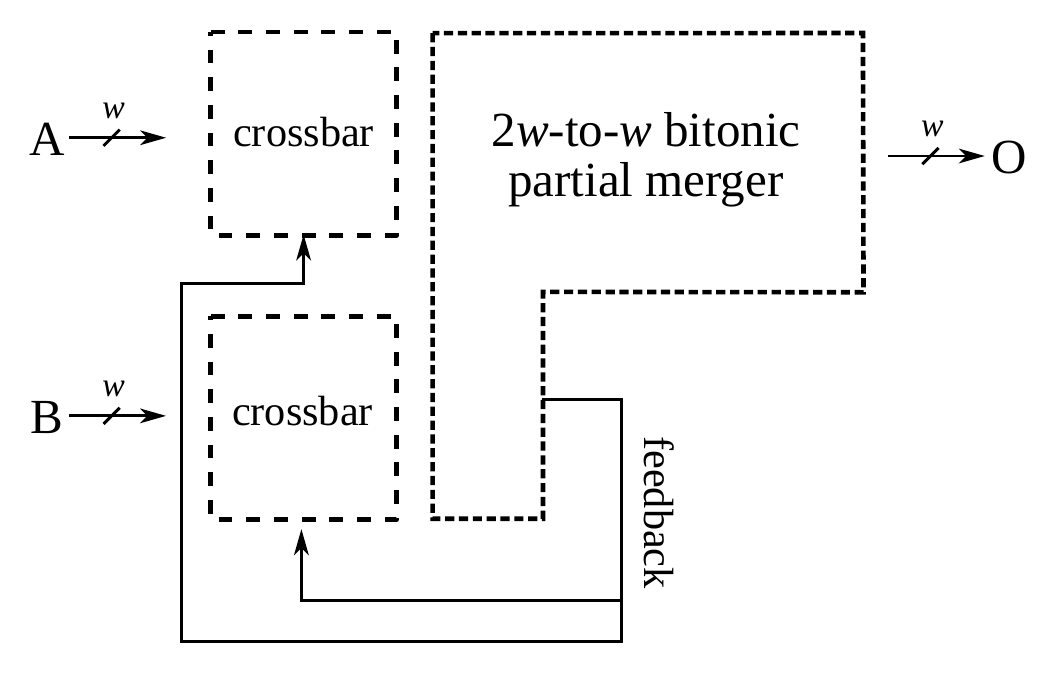}
\caption{High-level view of the merger used in PMT \cite{pmt}}
\label{songd}
\end{figure}

In 2016, Song et al. built the parallel merge tree (PMT \cite{pmt}) with 2\(w\)-to-\(w\) bitonic partial mergers. In figure \ref{songd}, we can see a high-level view of the merge block. This merger works as follows: two input queues, A and B, output 0 to \(w\) elements each per cycle, according to how many made it in the last result of \(w\) elements. This is known from just the first stage of the 2\(w\)-to-\(w\) bitonic partial merger and is used as a feedback to select the amount of elements to be dequeued from A and B. Since some elements remain from the previous cycle, each input of the bitonic partial merger block needs to be rotated by an offset equal to the number of dequeued elements (so far). This is done to ensure that the bitonic partial merger gets sorted inputs. %The problem with this approach is that the crossbars implementing the barrel shifters create a critical path that increases with \(w\). This leads to lower frequency designs and it does not scale very well \cite{mashimo2017high}. %
However, the crossbars implementing the barrel shifters create a critical path that increases with \(w\), leading to low frequency designs and it does not scale well \cite{mashimo2017high}.

At some point, most of the attention was drawn on removing the expensive feedback length that existed in traditional merger designs \cite{casper, pmt}, that prevented scalability in terms of operating frequency for an increased degree of parallelism (\(w\)) (SHMS \cite{mashimo2017high}, MMS \cite{mms}, VMS \cite{vms}).

In 2017, Mashimo et al. \cite{mashimo2017high} proposed a lower-latency-feedback architecture, SHMS, to solve the long critical path problem with the previous approaches. While achieving much higher frequencies, as high as 3.14 times more than PMT \cite{pmt} for 32 different input queues (also focusing on multiple inputs), the register utilisation was much higher (7.26 times more registers than PMT for 32 inputs). %
This does not scale well, not from the long feedback datapath length, but due to the high register utilisation.

\begin{figure}[h!]
\centering
\includegraphics[width=0.43\textwidth, trim=0  20 0 30 ]{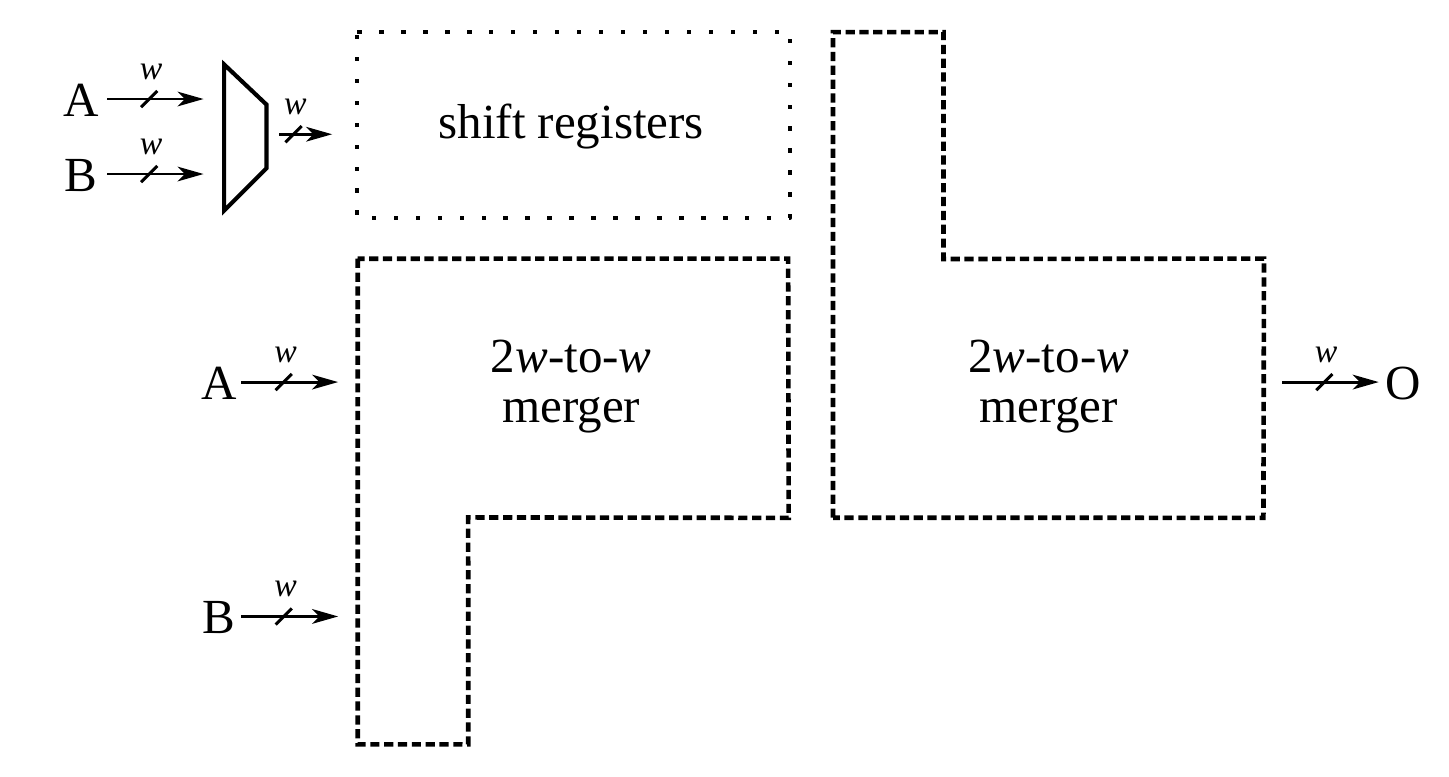}
\caption{High-level view of the mergers used in MMS \cite{mms} and VMS \cite{vms}}
\label{mmsd}
\end{figure}

Then, Saitoh et al. \cite{mms} proposed a feedback-less architecture, MMS, to increase the performance and scalability of the merge operator. MMS uses two 2\(w\)-to-\(w\) bitonic partial merge blocks along with shift registers and an extra comparator and multiplexer. VMS \cite{vms} is a variation of MMS that is based on odd-even mergers instead, but also focuses on improving the tie-record workaround (see section \ref{flimscomp}). In figure \ref{mmsd}, we can see the high-level view of these designs. %

\begin{figure}[h!]
\centering
\includegraphics[width=0.5\textwidth, trim=0 0 0 0]{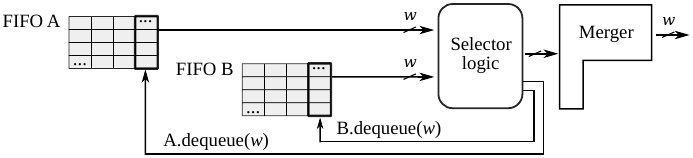}
\caption{Dequeue architecture used in WMS \cite{ehms}}
\label{ehmsf}
\end{figure}

Finally, FLiMS \cite{flims} and WMS/EHMS \cite{ehms} offered further improvements by focusing on efficiency, for minimising the required hardware resources, usually with a subsequent improvement in operating frequency. Essentially, WMS is an optimisation of MMS \cite{mms} (and VMS \cite{vms}), because it fuses the two 2\(w\)-to-\(w\) partial mergers into one bigger merger block, and also eliminates the need for additional shift registers. Figure \ref{ehmsf} shows how a single merger is used in WMS%, with a very close resemblance to the other approaches in MMS, VMS and EHMS.
, closely resembling the other approaches in MMS, VMS and EHMS.

EHMSP \cite{ehmsp}, was then introduced as a potential successor to EHMS \cite{ehms}. EHMS and EHMSP try to move some complexity to the selector stage for lower resource utilisation at the expense of a lower operating frequency. EHMSP specifically is not considered here for comparison, as its resource utilisation is close to EHMS, but with a further performance overhead due to the increased complexity of its selector stage, worsening its critical path \cite{ehmsp}. Also, in contrast to the other high-throughput mergers, EHMSP is for \(w\) values not in powers of 2, making it less versatile.

\section{A novel 2-way high-throughput merger}

FLiMS is a novel 2-way high-throughput merger %
that only uses a single 2\(w\)-to-\(w\) bitonic partial merger.  It merges 2 sorted inputs with high throughput without the use of barrel shifters or shift registers. 

\begin{figure}[h!]
\centering
\includegraphics[width=0.5\textwidth, trim=0 7 0 7]{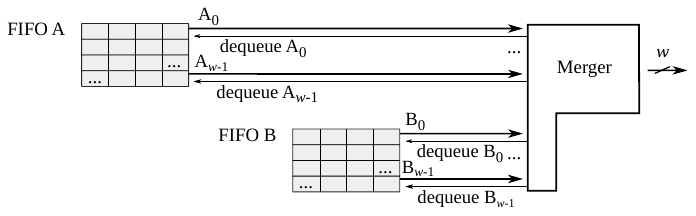}
\caption{Dequeue architecture of FLiMS \cite{flims}}\label{prarch}
\end{figure}

In figure \ref{prarch}, we can see a high-level visualisation of the proposed parallel merger. %
In contrast to previous works, FLiMS is shown to dequeue from the banks on an individual basis, rather than in batches of \(w\) elements, due to the decentralised selector logic.

\begin{figure}[h!]
\centering
\includegraphics[width=0.5\textwidth, trim=4 12 -4 5 ]{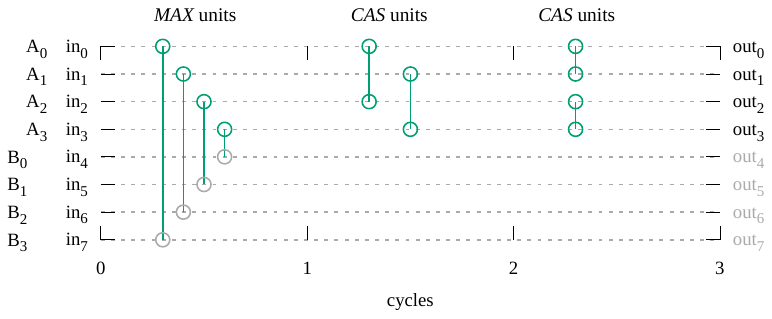}
\caption{FLiMS in low-level: \emph{MAX} selector stage plus a \emph{CAS} network (butterfly topology), \(w=4\).%
}\label{prarch2}
\end{figure}

In figure \ref{prarch2}, we see a lower-level representation, where the green circle pairs are the compare-and-swap units, with the exception of the first pipeline stage, which has \(w\) pairs of one green and one grey circle, representing the \emph{MAX} units (selector stage). %
If we ignore the modification of the first stage to \emph{MAX} units, this topology is the bitonic partial merger \cite{farmahini2012modular}, and it would produce the top \(w\) out of two sorted lists of \(w\) elements, and is a subset of the bitonic merger (see figure \ref{bits}).

The proposed parallel algorithm can be broken down into 3 segments: a \emph{selector stage} for handling the input from multiple banks/queues, the pipelined \emph{bitonic partial merger} (minus the first stage) and finally the \emph{output logic}.

\subsection{Selector stage} After the data are written into the BRAM banks (stored with a round-robin priority), a set of independent entities (nodes) are responsible for controlling the input to the merger. These \(w\) entities respond to the same clock and their behaviour can be described by a distributed algorithm. 

Let \(A\) and \(B\) be the input FIFO queues containing the two sorted lists to be merged. Let \(A_0\), ..., \(A_{w-1}\) and \(B_0\), ..., \(B_{w-1}\) be the queues corresponding to the respective banks for A and B. Each of the entities \(\mathit{MAX}_0\), \(\mathit{MAX}_1\), ..., \(\mathit{MAX}_{w-1}\) have as input the pair of queue heads (\(a_0\), \(b_{w-1}\)), (\(a_1\), \(b_{w-2}\)), ..., (\(a_{w-1}\), \(b_0\)) respectively, with \(a_i\) being the head of \(A_i\) and \(b_i\) being the head of \(B_i\). Each of these entities outputs one number per cycle when both inputs are valid. 

Each entity \(\mathit{MAX}_i\) has the data registers \(cA_i\) and \(cB_i\) to store the last heads that were dequeued from banks \(A_i\) and \(B_{w-1-i}\) respectively. It also has a register for the sorting network input (\(in_i\)). On each cycle with valid input, if \(cA_i\) $>$ \(cB_i\), it means that \(cA_i\) will make it into the top \(w\) in the result of the pipeline and therefore \(cA_i\) is copied into \(in_i\). In this case, \(cA_i\) is replaced by the head \(a_i\), which is dequeued from \(A_i\), but \(cB_i\) will remain unchanged, since it will need to be compared again in the next cycle, being in the lower \(w\). The equivalent logic goes for the case when \(cA_i\) $\leq$ \(cB_i\).

In algorithm \ref{alg1}, we can see the pseudocode for the distributed algorithm. Collectively, this algorithm  replaces the first stage of the partial bitonic merge (half-cleaner), with \emph{MAX} units instead of \emph{CAS} units. It selects the current top \(w\) on each cycle and inserts them into pipeline registers for the rest of the \emph{CAS} network to sort and produce the correct \(w\)-sized chunk of output.

\begin{algorithm}[h!]
\small
\LinesNumbered
\SetSideCommentRight
\SetKwComment{Comment}{$\triangleright$\ }{}
int \(i\)\Comment*[r]{\(i\) is the entity tag}
reg \(cA_i\), \(cB_i\), \(in_i\)\Comment*[r]{registers of data width}
 \While{forever}{
 receive (positive clock edge);\\
\eIf {\(cA_i\)$>$\(cB_i\)}{
	\(in_i\) $\leftarrow$ \(cA_i\);\\
	\(cA_i\) $\leftarrow$ dequeue(\(a_i\));\\
}{
	\(in_i\) $\leftarrow$ \(cB_i\);\\
	\(cB_i\) $\leftarrow$ dequeue(\(b_{w-1-i}\));\\
}  
  
 }
\caption{\(\mathit{MAX}_{i}\) unit pseudocode}\label{alg1}
\end{algorithm}

There may be a need for some extra logic required to correctly handle the ending of the input queues, but it is omitted here for the sake of simplicity and portability to different architectures, as it is relatively trivial to construct. %
For example, when sorting natural numbers in descending order, the value 0 can be passed afterwards to handle the ending without additional dedicated logic.

\begin{table*}[h!] 
  \caption{Merging two descending sequences: Example execution for \(w\)=4 and random sorted lists A and B.\(^1\) } 
\label{tab10} %
\vspace{-1em}
\footnotesize
\centering
  
\begin{tabular}{c}
\\
\end{tabular}
 
\begin{tabular} {c|l l c c r }
\hline

Cycle&Input A&Input B& cA& cB&Output \emph{(after the pipeline delay)}\\
\hline
&&&&&\\

0&3 3 4 5 11 16 17 26 26 29 &0 7 8 9 12 15 18 19 21 22 &&&\\
1&3 3 4 5 11 16 &0 7 8 9 12 15 &17 \underline{26} \underline{26} \underline{29} &\underline{22} 21 19 18 &22 26 26 29 \\
2&3 3 4 &0 7 8 9 12 &\underline{17} { }5 11 16 &15 \underline{21} \underline{19} \underline{18} &17 18 19 21 22 26 26 29 \\
3&3 3 &0 7 &{ }{ }4 { }5 \underline{11} \underline{16} &\underline{15} \underline{12} { }9 { }8 &11 12 15 16 17 18 19 21 22 26 26 29 \\
4& & &{ }4 { }\underline{5} { }3 { }3 &{ }\underline{7} { }0 { }\underline{9} { }\underline{8} &5 7 8 9 11 12 15 16 17 18 19 21 22 26 26 29 \\
5& & &{ }\underline{4}{ } { }{ }\space { }\underline{3} { }\underline{3} &  { }\space{} { }\underline{0} { }{ } { }{ } &0 3 3 4 5 7 8 9 11 12 15 16 17 18 19 21 22 26 26 29 \\
\hline
\multicolumn{6}{r}{}\\[-0.15cm]
\multicolumn{6}{r}{\(^1\)\emph{see \url{http://philippos.info/sort_visual} for an online visualisation.}}\\
\end{tabular}

\begin{tabular}{c}
\\
\end{tabular}
\vspace{-1.5em}
\end{table*}

\subsection{\emph{CAS} network}

The compare-and-swap (\emph{CAS}) network of FLiMS is responsible for sorting the top-\(w\) result of the selector stage. It is a partial bitonic merger minus the first stage, or in other words, a butterfly network. %
It is not considered a sorting network on its own and does not sort arbitrary number sequences.  For its input though, it behaves correctly and its output is always sorted (see proof \ref{proof1}).

\subsection{Output logic} %
On each cycle, if the output of the partial bitonic merger is marked as valid, it is written down as a \(w\)-sized chunk of the result, such as in \(w\) output banks containing implementing an output queue \(O\). %

One observation is that when there is valid output per cycle, it produces exactly \(w\) elements, as with other 2-way mergers. This is useful for easing synchronisation when embedding into a merge tree \cite{fsorter}. %

The sets of registers cA and cB are not visible in the first (selector) pipeline stage of figure \ref{prarch2}. The notion of those registers is optional because they can also be considered the current sets of heads of the banked queues A and B. Though, it is sometimes convenient to use cA and cB, such as when the input queues are block RAM sections, where a read register is already present for reading each memory. 

Table \ref{tab10} presents an example execution for \(w\)=4.

\section{Additional functionality}\label{afunc}
 
This section presents variations of FLiMS, which can be used to increase its applicability or performance, according to the distribution of the data and the requirements of the sorting problem and platform.

\subsection{Skewness optimisation} \label{sdo}

FLiMS can be used to build parallel merge trees that merge many input lists hierarchically in a single pass. Parallel merge trees can suffer from rate mismatch that occurs when the input data distribution leads to underutilisation of certain mergers, resulting in reduced throughput \cite{pmt, leaf, christopher}. 

While the memory access throughput is a matter of the memory system, it might be allowed for the accelerator to receive the data from each of the input lists A and B with a fixed bandwidth, less than \(w\), such as in a PMT. A contributor to rate mismatch %
is when there are a lot of duplicates in the input (skewed datasets).  %
When the data are skewed, the merger only dequeues from one of the input queues for long periods of time. This results in stalled cycles from underutilising the aggregate bandwidth of the queues.

PMT \cite{pmt} proposes a simple solution which causes the merger blocks to fetch from both inputs at a similar rate %
when there are duplicates. 
This ensures that the input queues are consumed with a similar throughput, that collectively balances the utilisation of the merge tree. However, PMT's mergers inherit the long feedback problem, which was addressed in subsequent works \cite{mms, vms, flims, ehms}. 

We propose the equivalent optimisation for FLiMS \cite{flims}%
, while keeping the decentralised nature of the selector stage. On duplicates, there is an ``oscillating'' effect at the \emph{MAX} units, which balances the dequeuing rate from the two groups of inputs. The code for the new selector units is illustrated in algorithm \ref{al2}. An 1-bit register called \(dir_{i}\) represents the input out of which the result was taken during the previous cycle, and is appended to the least significant bit in the comparison, to enforce a sort priority on equal values.

\begin{algorithm}[!h]
\small
\SetSideCommentRight
\SetKwComment{Comment}{$\triangleright$\ }{}
int \(i\)\Comment*[r]{\(i\) is the entity tag}
reg \(cA_i\), \(cB_i\), \(in_i\)\Comment*[r]{registers of data width}
\textcolor{blue}{reg \(dir_{i}\)\Comment*[r]{1-bit register}}
 \While{forever}{
 receive (positive clock edge);\\
\eIf {\(\textcolor{blue}{\{}cA_i\textcolor{blue}{, dir_{i}\}}\)$>$\(\textcolor{blue}{\{}cB_i\textcolor{blue}{, !dir_{i}\}}\)}{
	\(in_i\) $\leftarrow$ \(cA_i\);\\
	\(cA_i\) $\leftarrow$ dequeue(\(A_i\));\\
	\textcolor{blue}{\(dir_{i}\) $\leftarrow$ \(0\);\\}
}{
	\(in_i\) $\leftarrow$ \(cB_i\);\\
	\(cB_i\) $\leftarrow$ dequeue(\(B_{w-1-i}\));\\
	\textcolor{blue}{\(dir_{i}\) $\leftarrow$ \(1\);\\}
}    
 }  
\caption{ Modified \({MAX}_i\) unit pseudocode for the skewness optimisation\label{al2}}
\end{algorithm}

\subsection{Stable merge}\label{stame}

In contrast to the skewness optimisation, stable sort may be desired instead. Stable sort is when the sorted output has the same order for duplicate values as they appear in the input. For implementing a stable merge sort, FLiMS would also need to be stable, i.e. to prioritise the duplicates of input A over the ones from input B, and keep their original order inside A and B accordingly. Such a modification cannot co-exist with the skewness optimisation, since the priority between the duplicates will be based on the input source.  

Originally, FLiMS is not stable, as it is partly-based on the bitonic sorter, which is not stable. Temporarily appending the input source (1 bit) and the port number (\(log_2(w)\) bits) to the MSB would be required to disambiguate between the original order of duplicates inside the \emph{CAS} network. As the order of the inputs inside \emph{MAX} units are naturally rotated by an offset, the port order is not enough to distinguish the order inside each batch containing duplicates. For this reason, a 2-bit value needs to be carried between the input source and the port number, that keeps count of the batch order. %
A single-bit counter would not be enough for distinguishing which of the two compared entries came first.  %
Algorithm \ref{al2m} shows the modifications required on the \emph{MAX} units to support stable merge.

\begin{algorithm}[!h]
\small
\LinesNumbered
\SetSideCommentRight
\SetKwComment{Comment}{$\triangleright$\ }{}
int \(i\)\Comment*[r]{\(i\) is the entity tag, \textcolor{blue}{\(log_2(w)\)-bits}}
reg \(cA_i\), \(cB_i\), \(in_i\)\Comment*[r]{registers of data width}
\textcolor{blue}{reg \(orderA_i=0, orderB_i=0\)\Comment*[r]{2-bit registers}}
 \While{forever}{
 receive (positive clock edge);\\
\eIf {\(cA_i\)$>$\(cB_i\) \textcolor{blue}{\(|| cA_i==cB_i\)}}{
	\(in_i\) $\leftarrow$ \textcolor{blue}{\(\{1,orderA_i,w-1-i,\)}\(cA_i\)\textcolor{blue}{\(\}\)};\\%
	\(cA_i\) $\leftarrow$ dequeue(\(A_i\));\\
	\textcolor{blue}{\(orderA_{i}\) $\leftarrow$ \(orderA_{i}-1\)};\\%
}{
	\(in_i\) $\leftarrow$ \textcolor{blue}{\(\{0, orderB_i, i, \)}\(cB_i\)\textcolor{blue}{\(\}\)};\\%
	\(cB_i\) $\leftarrow$ dequeue(\(B_{w-1-i}\));\\
	\textcolor{blue}{\(orderB_{i}\) $\leftarrow$ \(orderB_{i}-1\)};\\%
}    
}  
\caption{ Modified \(\mathit{MAX}_i\) unit pseudocode for implementing stable merge in descending order \label{al2m}}
\end{algorithm}
Additionally, the \emph{CAS} units also need to be modified to correctly prioritise the case where the 2-bit order is ``00'' against ``11'', as all other combinations (same values or other pairs having a difference of one) would correctly represent the original order priorities. %
The general idea of this approach is to emulate appending the original input order to the MSB of the data, but with a steady and low number of bits for merging arbitrarily long input. The order field can be seen as the last few bits of the input order.

\subsection{Dequeuing whole rows (FLiMSj) }\label{flimsj}

One potential advantage of the majority of the related work \cite{casper,mms,vms,ehms,ehmsp} over FLiMS is that they dequeue whole rows of \(w\) elements (or \(w/2\) for EHMS \cite{ehms}) from the inputs by default (see figure \ref{ehmsf}). This reduces the number of dequeue signals, and can also be more efficient in special cases, such as when reading narrow data from wider memories. %

A relatively efficient buffering arrangement can unify the dequeue signals for FLiMS as well. %
This is possible because the FIFOs in FLiMS are collectively consumed in round-robin fashion and at no point two FIFO indexes of the same input differ by more than one. This %essentially 
means that a set of registers for buffering the next queue heads is enough for dequeuing an element batch at the right time, while providing full bandwidth for the respective input. %We notice that 
This can be achieved with a single set of \(w\) registers for both inputs. 

\begin{figure}[h!]
\centering
\includegraphics[width=0.5\textwidth, trim=0 10 0 -7 ]{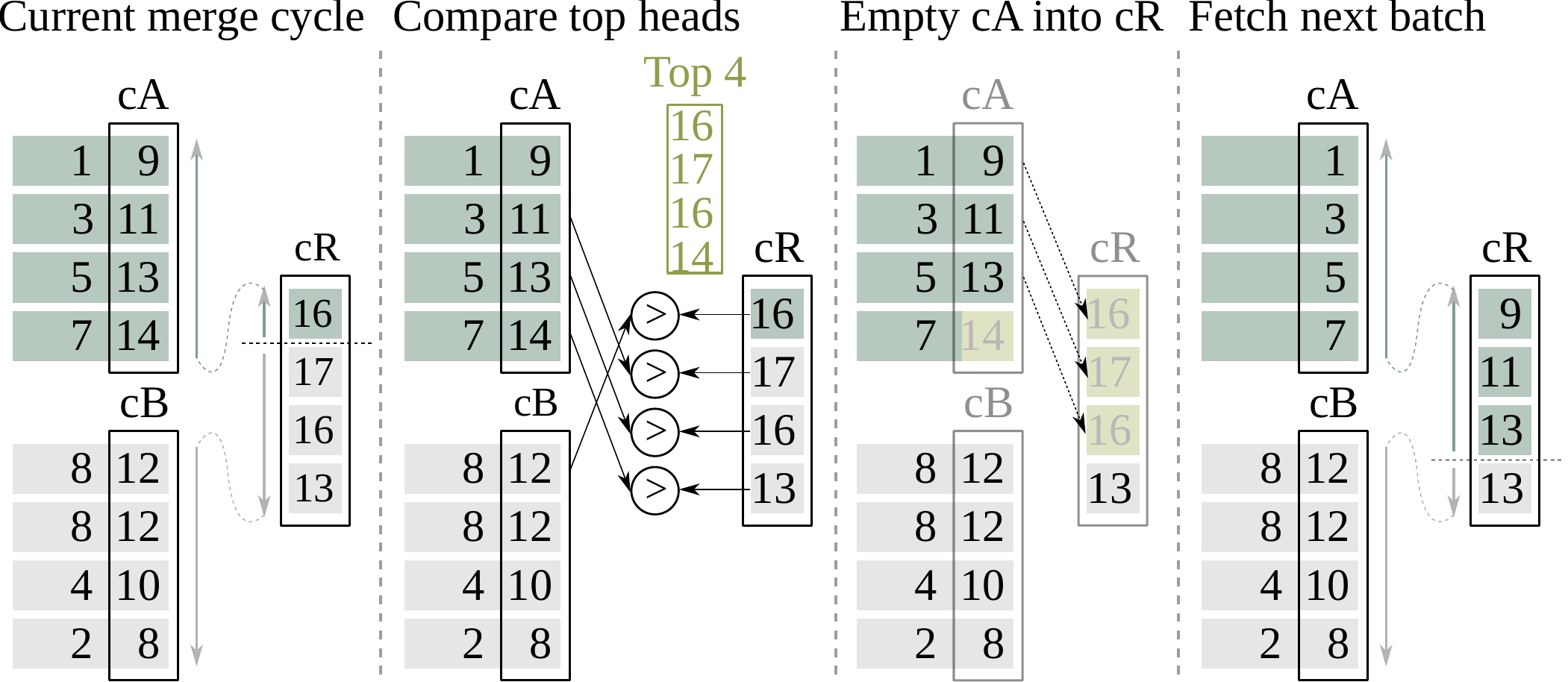}
\caption{FLIMSj merging example for dequeuing whole rows, \(w=4\).%
}\label{flimsjfig}
\end{figure}

Figure \ref{flimsjfig} introduces the related modification to the \emph{MAX} units using a merging example for \(w=4\). 
The general idea is that the top \(2w\) to \(w\) elements can be stored in a set of \(w\) registers (\(cR\)) after every selection iteration (cycle), while maintaining their natural rotation order. This results in at least \(w\) available elements per input, when combined with the current heads in \(cA\) and \(cB\), while still eliminating the need for a rotation and a growing feedback. Algorithm \ref{algfj} describes this approach in more detail.

\begin{algorithm}[t!]
\small
\LinesNumbered
\SetSideCommentRight
\SetKwComment{Comment}{$\triangleright$\ }{}
int \(i\)\Comment*[r]{\(i\) is the entity tag}
reg \(cA_i\), \(cB_i\)\textcolor{blue}{, \(cR_i\)}, \(in_i\)\Comment*[r]{registers of data width}
\textcolor{blue}{reg \(dir_i\), \(src_i\)\Comment*[r]{1-bit registers}}
 \While{forever}{%
 receive (positive clock edge);\\
 \textcolor{blue}{ \Comment*[h]{(use the correct head based on \(src_i\))}\\}
\eIf {\textcolor{blue}{\((src_i\ ?\ \)}\(cA_i\) \textcolor{blue}{\(: cR_i)\)} $>$ \textcolor{blue}{\((src_i\ ?\ cR_i :\ \)}\( cB_i\)\textcolor{blue}{\()\)}}{
	\(in_i\) $\leftarrow$ \textcolor{blue}{\((src_i\ ?\ \)}\(cA_i\) \textcolor{blue}{\(: cR_i)\)};	
	\textcolor{blue}{\\\(dir_{i}\) $\leftarrow$ \(0\);\\}
}{
	\(in_i\) $\leftarrow$ \textcolor{blue}{\((src_i\ ?\ cR_i :\ \)}\( cB_i\)\textcolor{blue}{\()\)};	
	\textcolor{blue}{\\\(dir_{i}\) $\leftarrow$ \(1\);\\}
}  
\textcolor{blue}{
sync(\(dir_i\))\Comment*[r]{(data dependency)}
\If{\(src_i==dir_i\)}{ %
\Comment*[f]{(the consumed element is from \(cR_i\))}\\
	\(src_i\) $\leftarrow$ \(dir_0\);\Comment*[f]{(next following \(\mathit{MAX}_0\))}\\
	\(cR_i\)  $\leftarrow$ \(dir_0\)\ ?\ \(cB_i : cA_i\);\\
  }  
  \Comment*[h]{Fetch the next batch collectively}\\
  \(dir_0\)\ ? $cB_i\leftarrow$ dequeue(\(b_{w-1-i}\)) : $cA_i\leftarrow$ dequeue(\(a_i\));\\
}
}
\caption{Modified \(\mathit{MAX}_{i}\) unit pseudocode for dequeuing whole rows from the inputs (FLiMSj)}\label{algfj}
\end{algorithm}

\section{Correctness}\label{corr}

This section provides proofs on some non-trivial parts of our proposed techniques. The correct operation of FLiMS and its skewness optimisation can be proven by induction. FLiMS can be used for merging lists in descending order, as well as for merging ascending lists with minor modifications (reversing all comparators, and reversing the order in the stable version). Here, we study FLiMS designs with data in descending order, but the proofs can be easily adapted %
for the other case, without loss of generality.

\subsection{Without additional functionality %
}\label{proof1}

In order to prove that the main design behaves as expected we will show that it is functionally equivalent to a more trivial merger implementation, where on each cycle the input comes sorted to a 2\(w\)-to-\(w\) bitonic partial merger, as with the merger used in PMT \cite{pmt}. We will show that (1) the selector stage algorithm always selects the top \(w\) out of the 2\(w\)-sized input and that (2) the butterfly network always sorts this top \(w\) list before writing the result in output.

(1) %
We denote as \(l_A, l_B\in\{0,1,...,w-1\}\) the naturally-occurring rotation offsets for input queues A and B respectively. Supposedly, the selector stage dequeues consecutive elements from each banked input.
Since A and B are written in a round robin fashion inside the banked memory, %there are different rotation offsets that would be required to read 
different rotation offsets would be required to read
a sorted set of \(w\) elements from each set of FIFOs, %at least to correspond to a design that rotates its inputs, as in PMT \cite{pmt}. 
similar to a design that rotates its inputs like PMT \cite{pmt}.
Note that the \emph{MAX} units in FLiMS receive elements from the corresponding banks without performing any additional 
rotation. %

\ul{Induction hypothesis:} On each cycle, if the selector stage has worked correctly on the previous cycle, it will load \(k\in \{0,1,...,w\}\) %
elements from A and \(w-k\) from B, collectively corresponding to the combined top \(w\) out of the available \(2w\) elements. %

\ul{Base step:} At cycle 0, the parallel merger behaves in the same way as the 2\(w\)-to-\(w\) bitonic partial merger, because the \(w\) inputs from each list are already sorted (\(l_A=0\) and \(l_B=0\)), corresponding to the first \(w\) elements from A and first \(w\) from B (stored in \(cA\) and \(cB\) registers respectively). The first stage of the 2\(w\)-to-\(w\) bitonic partial merger is known as a half-cleaner, and produces the top \(w\) out of the \(2w\) inputs, in a bitonic sequence. %

\ul{Induction step:} 
Given that the selector stage worked correctly in the previous cycle, \(k' \in\{0,1,...,w\}\)  and \(w-k'\) are the amounts of sorted elements dequeued last from A and B respectively. 

On each cycle, \(l_A\) and \(l_B\) will be updated according to the number of dequeued heads, as the starting positions shall succeed the dequeued elements. That is, \(l_A=(l_A'+k')\bmod w\) for A, and \(l_B=(l_B'+(w-k')) \bmod w\) for B, where \(l_A'\), \(l_B'\) the offsets of the previous cycle. %
Therefore,

\vspace{-1em}
\begin{align*} 
&\ l_B=(l_B'+(w-k')) \bmod w\\
\Rightarrow &\ l_B=(l_B'+(w-(l_A-l_A'))) \bmod w\\
\Rightarrow &\ l_B=(l_B'-l_A+l_A') \bmod w\\
\Rightarrow &\ (l_B+l_A)\bmod w=(l_B'+l_A') \bmod w
\end{align*}

As cycle 0 assumes %
\(l_A'=0\) and \(l_B'=0\) (due to the input already being aligned correctly in the banks), %
and that the induction hypothesis is assumed correct for all previous cycles, it always holds that \((l_B'+l_A')\bmod w=0\). %
Since \(l_B'=l_B' \bmod w\), 
%
%it holds that 
\(l_B = (w-l_A) \bmod w\) for every cycle.

In order for the selector stage to produce the top \(w\) out of the current \(2w\) elements, the comparisons that need to be made are between all pairs in \(\{(Ta_i, Tb_{w-1-i})| \forall i \in\{0,1,...,w-1\}\}\)%
, where \(Ta_0\), ..., \(Ta_{w-1}\) are the current top \(w\) elements in A and \(Tb_0\), ..., \(Tb_{w-1}\) are the current top \(w\) elements in B, sorted in descending order. These comparisons are required in order to emulate the first stage (half-cleaner) of the bitonic partial merger.

The \emph{MAX} units have the same topology as the bitonic (partial) merger (as seen in figure \ref{bits}). We notice that whatever the rotation combination from \(l_A\) and \(l_B\) is, the correct comparisons will be made, as: 

\vspace{-1em}
\begin{align*} 
&\ (Ta_{(i+l_A) \bmod w}, \ Tb_{((w-1-i)+l_B) \bmod w}) \\ 
\equiv &\ (Ta_{(i+l_A) \bmod w}, \ { }Tb_{((w-1-i)+(w-l_A)) \bmod w}) \\ 
\equiv &\ (Ta_{(i+l_A) \bmod w},\ { } Tb_{(w-1-i-l_A) \bmod w}) \\
\equiv &\ (Ta_{(i+l_A) \bmod w},\  { }Tb_{(w-1-(i+l_A)) \bmod w}).
\end{align*}

(2) The sorting network receives the correct top-\(w\) output from (1) and the task is to sort it. % correctly. 
Originally, this structure is supposed to sort bitonic sequences of size \(w\) \cite{batcher}. The output of the distributed algorithm block is a rotated bitonic sequence, as we saw that the comparisons will be rotated by \(l_A\). A rotated bitonic sequence is also a bitonic sequence \cite{zachmann2013adaptive}, therefore the input for the sorting network has the correct property. This completes the proof.

\subsection{Including the skewness optimisation}

\begin{table*}[h!]  
\caption{Comparing high-throughput 2-way mergers}\label{flimscompt}
  \centering
  \begin{tabular}{  c | c | c | c | c | c | c}
    \multirow{2}{*}{2-way merger} & Feedback & \multirow{2}{*}{Latency} & \multirow{2}{*}{Number of comparators} & \multirow{2}{*}{H/W modules}&Merger &Tie-record\\ 
    &length&&&&topology&challenge\\\hline
    &&&&&\\
     basic \cite{casper, simd2008}
   & \(\log_2(w)+2\)& \(\log_2(w)+2\) &\(w+w \log_2 (w)\)&\(1\times\)\emph{2w-to-2w} merger%
&bitonic&no \\ %

     from PMT \cite{pmt}
   & \(\log_2(w)+1\)& \(2 \log_2(w)+1\) &\(w+\frac{1}{2}w \log_2 (w)\)&\thead{\(1\times\)\emph{2w-to-w} merger \\ \& 2 barrel shifters%
   }&bitonic&no \\ %

      MMS \cite{mms}
    & 1 &  \(2  \log_2(w) + 3\)  &\(2 w+w \log_2 (w)\)+1& \thead{\(2\times\)\emph{2w-to-w} mergers \\ \& shift registers}&bitonic&yes\\ %

      VMS \cite{vms}%
    & 1 &  \(2  \log_2(w) + 3\)  &\(2 w+w \log_2 (w)\)+1& \thead{\(2\times\)\emph{2w-to-w} mergers \\ \& shift registers}&odd-even&yes\\ %

      WMS \cite{ehms, ehmsp}
    & 1 & \(\log_2(w) + 3\)  &\(3w+\frac{1}{2}w \log_2 (w)\)&  \thead{\(1\times\)\emph{3w-to-w} merger\\}&odd-even&yes \\ %

      EHMS \cite{ehms, ehmsp}
    & 1 & \(\log_2(w) + 3\) &\(\frac{5}{2}w+\frac{1}{2}w \log_2 (w)+2\)&  \thead{\(1\times\)\emph{2.5w-to-w} merger\\}&odd-even&yes \\ %
    &&&&&\\
      FLiMS \cite{flims}
    & 1 & \(\log_2(w)+1\) &\(w+\frac{1}{2}w \log_2 (w)\)&\(1\times\)\emph{2w-to-w} merger&bitonic&no\\ %
    &&&&&\\
      FLiMSj
    & 1 & \(\log_2(w)+2\) &\(w+\frac{1}{2}w \log_2 (w)\)&\(1\times\)\emph{2w-to-w} merger&bitonic&no\\ 

  \end{tabular}
  \small
    \begin{tabular}{  l }
  \\
  \end{tabular}
%
%\vspace{-0.5em}
\end{table*}

To prove that FLiMS continues to sort correctly, the selector stage must be shown to still produce a bitonic sequence \cite{flims} (up to one local maximum and up to one local minimum). 

The bitonic sequence property needs to apply also on the order of each input when there are duplicates. Therefore, we need to show that there will still be up to one local maximum and up to one local minimum in the bitonic sequence, even though there might be multiple additional entries with a value equal to the minimum or maximum. %

The original order between consecutive duplicates in the same input is used to correctly prioritise duplicates, as the input lists are considered already sorted. Being consistent about the original order ensures dequeuing consecutive entries from A and B, keeping the integrity of the input data, as the data are stored in round-robin fashion in banks. 

On each cycle, each %of the \(w\) 
\(\mathit{MAX}_i\) units compares \(Ta_j\) to \(Tb_{w-1-j}\), where \(i\) is a rotation of \(j\) by a common offset \(o \in \{0, 1, ..., w-1\}\) (\(o=l_A=(w-l_B) \bmod w\)), and \(Ta\) and \(Tb\) represent the remaining elements of the input queues in descending order. %On the basic version, 
This emulates a half-cleaner that selects (and dequeues) a total of the greatest \(w\) elements from a total of \(2w\) elements, generating a bitonic sequence.

\ul{Induction hypothesis:} On each cycle, if the skewness-optimised selector stage worked correctly on the previous cycle, it will produce a bitonic sequence from \(k\) consecutive elements from A and \(w-k\) consecutive elements from B, where \(k\in \{0,1,...,w\}\).

\ul{Base step:} At cycle 0, the common offset is zero (\(o=0\)), as the inputs are properly aligned inside the input FIFOs, and no element has been dequeued yet. Each \(\mathit{MAX}_i\) unit has an initial value of 0 stored in its \(dir_{i}\) register. This means that on the event of comparing duplicates, it will behave as in the non-optimised case, where a single source (B) is preferred for duplicates, as the comparison of line 6 of algorithm \ref{al2} is now equivalent to line 5 of algorithm \ref{alg1} ( ``if \(cA_i>cB_i\) then'' ), which is already proven to produce a bitonic sequence from the proof of section \ref{proof1}.

\ul{Induction step:} The skewness optimisation modification only takes effect where there are duplicates, i.e. \(\exists h\in \{0, 1, ..., w-1\}: Ta_{h}=Tb_{w-1-h}\). In such a case, we notice that this only happens consecutively (including wrap-arounds from the natural rotation) and for the minimum value of the output, as \((Ta_0, Ta_1, ..., Ta_{w-1})\) is monotonically decreasing and \((Tb_{w-1}, Tb_{w-2}, ..., Tb_0)\) is monotonically increasing. As a consequence, the position of the minimum (split) in the bitonic sequence can be at the start, end or between this region of duplicates. 

Given that FLiMS worked correctly on the last cycle, the \(dir_{i}\) registers correspond to the last half-cleaner decisions, which will be of the form \(\{1\}^{w-k}\{0\}^{k}, k\in \{0,1,...,w\}\), after considering the offset \(o\). Ones and zeros appear consecutively, since only consecutive elements are dequeued from each input list on each cycle. Also, the sequence of \(dir_{i}\) registers starts from 1, when we consider the current rotation offset. This is because the naturally-occurring rotation offset is updated according to the last position dequeued from \(A\), the next of which corresponds to the first from list \(B\), that yields the ``first'' 1 in the \(dir_{i}\) sequence.  

Therefore, the region of duplicates will be a sublist of the expression  \(\{1\}^{w-k}\{0\}^{k}\), with its 1s and 0s replaced by consecutive duplicates from \(A\) and \(B\) respectively. As a result, there will be up to one local minimum (split) in this region, and therefore up to one local minimum in the entire half-cleaner result, which consists a bitonic sequence.

\section{Comparison with the related work}\label{flimscomp}

Table \ref{flimscompt} compares FLiMS to the related work, according to different terms contributing to the resource utilisation and efficiency of the design. FLiMS uses the least amount of resources by only requiring a single 2\(w\)-to-\(w\) bitonic partial merger. It has a single-stage feedback latency, making it ``feedback-less'', and has the least amount of latency, which is \(log_2(2w)\). FLiMSj of section \ref{flimsj} is also added to the table, even though its only modification is in its \emph{MAX} units, which results in one more cycle of pipeline latency. %

The first two entries \cite{casper, pmt} have a feedback consuming multiple stages, such as from the additional \(\log_2(w)\) stages required to implement the barrel shifters before the inputs in the mergers of PMT \cite{pmt}. Hence, an increased number of inputs has a scalability problem, as a pipelined implementation of the feedback would reduce the throughput. Alternatively, squeezing the increasing logic into a single pipeline stage (consuming a single cycle) can heavily impact the operating frequency \cite{mms}. %

MMS \cite{mms} and VMS \cite{vms} were the first solutions that provided a practical solution for the feedback problem with a relatively low resource utilisation. Their approach was to use either two 2\(w\)-to-\(w\) bitonic partial mergers (MMS) or two 2\(w\)-to-\(w\) odd-even mergers (VMS). Both of those topologies have \(log_2(2w)\) pipeline stages and are relatively similar.  They are from the last \(log_2\) steps from the bitonic sorter and odd-even merge sort respectively \cite{batcher}.

WMS and EHMS \cite{ehms} on the other hand, achieve to use a single feedback-less merger, as with FLiMS, but this merger is for double the inputs (plus optimisations), totalling to one more pipeline stage. 

\begin{figure}[h!]
\centering
\includegraphics[width=0.48\textwidth, trim=0 8 0 0]{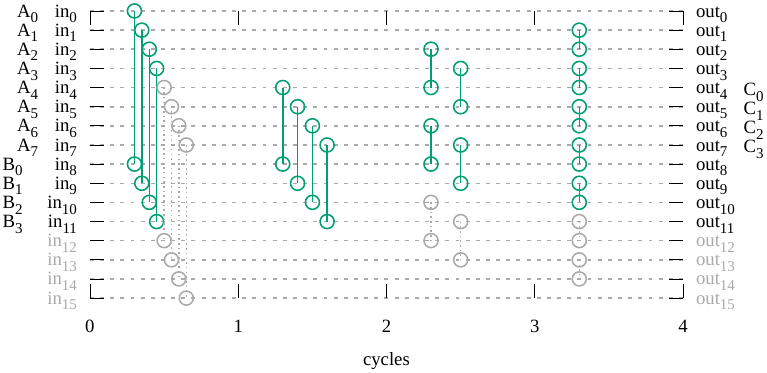}
\caption{Merger used in WMS and EHMS \cite{ehms}}\label{fi2bwm}
\end{figure}

Figure \ref{fi2bwm} shows how a \emph{4w-to-4w} merger from odd-even mergesort is adopted to implement \emph{3w-to-3w}, by pruning \emph{CAS} units. There is, though, additional pruning as the output is only \(w\) elements. EHMS uses the same merger, but the first \(w/2\) values of the input are not used, resulting in fewer comparisons. The resulting amount of comparators for each approach is shown in the table \ref{flimscompt}, and the formulas mainly derive from Cullen numbers \cite{cullen}, and are validated by using yosys through synthesising the Verilog implementations of the evaluation section \ref{fpgaimplem}.

The pipeline length (latency) also impacts the resource utilisation, as certain values need to be propagated for longer through the pipeline registers. The merger of WMS and EHMS uses one more cycle than FLiMS, as it is an optimised merge block for double the inputs (as seen in figures \ref{prarch2} and \ref{fi2bwm}). Note that WMS and EHMS propose an optimisation to reduce the pipeline registers by fusing some \emph{CAS} units, but it is ignored in this comparison, as it can be explored separately for all mergers. All mergers other than FLiMS and in PMT \cite{pmt}, have a separate single-cycle selector stage, contributing to one more pipeline stage, while in FLiMS it is integrated in the modified bitonic partial merger.

One challenge with the mergers MMS \cite{mms}, VMS \cite{vms}, WMS and EHMS \cite{ehms} is that, if there are duplicate values being compared, the output can be corrupted, also known as the \emph{tie-record issue}. Specifically, this is a problem in key-value pairs, where only the key is compared, and the integrity of the values can be lost where there are duplicate keys. The available workarounds of the related works vary in complexity (not presented). In FLiMS, this is not the case, as the selector stage decides for the top \(w\) result to propagate through the output immediately. In the other approaches, finding the top \(w\) is done more indirectly, relying on two orders, and the problem arises due to the odd-even merge and bitonic sort topologies not implementing stable sort.

\section{FPGA implementation}\label{fpgaimplem}

In order to evaluate FLiMS on FPGAs, we compare its resource utilisation and maximal operating frequency with the latest alternatives WMS and EHMS \cite{ehms}.  Our comparison includes FLiMSj for including a more direct competitor to the state-of-the-art WMS and EHMS mergers, when dequeuing whole rows is required (see section \ref{flimsj}). All generator scripts are implemented from scratch and produce Verilog code for each of the compared mergers for a given degree of parallelism \(w\) and data width. This experiment uses 64-bit mergers and targets the Xilinx Alveo U280 board. %

The generated designs work as simple AXI peripherals, that read already sorted data stored in distributed memory (on-chip) and write back also to distributed memory. The host places the sorted sublists and reads the merged result for validation purposes. For every different value of \(w\), the FIFOs are only 2 elements deep (totalling \(4w\) elements for input and output) to eliminate the differences between different merger designs to only their core logic. (Though, the bitstreams generated for validation had longer queues).

In order to simplify the comparison, the fusion of some pairs of compare-and-swap units (\emph{CAS}) in WMS and EHMS \cite{ehms} is not followed in this evaluation. This omitted optimisation could be explored separately, as with removing pipeline registers \cite{pswitch}, and does not directly relate to the main structure of the mergers. %
Additionally, the tie-record workarounds of WMS and EHMS have not been taken into consideration, even while FLiMS does not suffer from the tie-record issue. Thus, unique input values are assumed, such as with timestamp information inside the 64-bit input, or no satellite/payload data (i.e. values in key-value pairs). %

\begin{table}[h!]  
\caption{Resource utilisation as AXI peripherals, as reported by Vivado}\label{vresources}
  \centering
  \setlength{\tabcolsep}{4pt}
  \begin{tabular}{  r | r r | r r | r r | r r}

&\multicolumn{2}{c|}{FLiMS}&\multicolumn{2}{c|}{FLiMSj}&\multicolumn{2}{c|}{WMS}&\multicolumn{2}{c}{EHMS}\\
\(w\)&kLUT&\textcolor{gr}{kFF}&kLUT&\textcolor{gr}{kFF}&kLUT&\textcolor{gr}{kFF}&kLUT&\textcolor{gr}{kFF}\\
\hline
&&&&&&\\
4&1.7&\textcolor{gr}{2.9}&2.5&\textcolor{gr}{3.2}&2.7&\textcolor{gr}{5.3}&3.1&\textcolor{gr}{4.8}\\
8&3.6&\textcolor{gr}{6.3}&5.1&\textcolor{gr}{6.8}&5.6&\textcolor{gr}{11.0}&6.2&\textcolor{gr}{10.3}\\
16&7.0&\textcolor{gr}{1.4}&10.6&\textcolor{gr}{14.6}&11.7&\textcolor{gr}{23.1}&13.0&\textcolor{gr}{21.6}\\
32&15.4&\textcolor{gr}{29.0}&20.9&\textcolor{gr}{31.2}&23.5&\textcolor{gr}{48.3}&26.7&\textcolor{gr}{45.3}\\
64&33.7&\textcolor{gr}{62.0}&45.0&\textcolor{gr}{66.4}&53.3&\textcolor{gr}{100.8}&57.9&\textcolor{gr}{94.6}\\
128&73.4&\textcolor{gr}{132.2}&96.1&\textcolor{gr}{140.8}&106.6&\textcolor{gr}{209.8}&120.4&\textcolor{gr}{197.5}\\
256&158.6&\textcolor{gr}{280.7}&208.6&\textcolor{gr}{297.9}&224.0&\textcolor{gr}{436.0}&252.2&\textcolor{gr}{411.4}\\
512&345.3&\textcolor{gr}{594.0}&436.2&\textcolor{gr}{628.4}&473.0&\textcolor{gr}{904.7}&525.3&\textcolor{gr}{855.6}\\
  \end{tabular}
  \small
    \begin{tabular}{  l }
  \\
  \end{tabular}
\end{table}

\begin{figure}[h!]
\centering
\includegraphics[width=0.4\textwidth, trim=0 7 0 0 ]{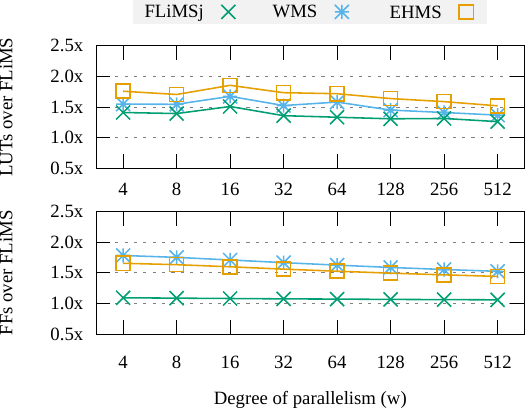}
\caption{Comparing resources of the state-of-the-art, FLiMSj and FLiMS}
\label{flimsortfall}
\end{figure}

Table \ref{vresources} includes the dataset on the obtained resource utilisation of look-up-tables (LUTs) and flip-flop registers (FF), as reported by Vivado 2020.1. Figure \ref{flimsortfall} is based on the same data and uses FLiMS as a baseline to emphasise on the overheads of the alternative approaches on resource utilisation.
As a conclusion, FLiMS uses the least amount of LUTs and FFs, while WMS and EHMS only differ marginally, as they are based on a similar merger. As expected \cite{ehmsp}, between WMS and EHMS, WMS wins in LUT utilisation, while EHMS wins in FF utilisation. Using the current implementation as an AXI peripheral, FLiMS is roughly about 1.5 to 2 times more hardware resource efficient. FLiMSj has almost the same FF utilisation as FLiMS, though in terms of LUTs it is about 1.3x more expensive than FLiMS, but always more resource efficient than WMS and EHMS.

\begin{figure}[h!]
\centering
\includegraphics[scale=0.79, trim=0 10 0 10 ]{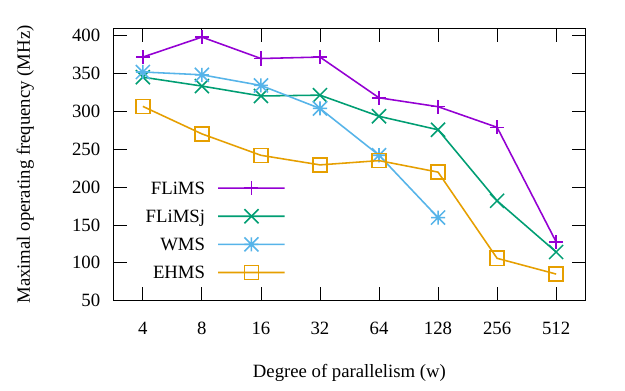}
\caption{Maximal operating frequency for FLiMS(j), WMS and EHMS}
\label{flimsortfall2}
\end{figure}

Finally, figure \ref{flimsortfall2} presents the comparison of the obtained maximal operating frequencies through the reported worst negative slack (WNS). Most datapoints used the default Vivado 2020.1 settings, though additional directives such as aggressive explore were used on some outliers or non-routable designs, especially for \(w\geq128\). Having such irregularity or small variations in the results are expected, as place-and-route is heuristic-based and becomes more challenging for larger designs. For WMS with \(w\geq256\), the additional tested directives did not help with routability. 

FLiMS has a considerable advantage over both WMS and EHMS, sometimes yielding more than double the operating frequency. %
FLiMSj has a small overhead over FLiMS, though WMS seems to marginally win for \(w\leq16\).
WMS is known to be better performing than EHMS, at the expense of additional hardware resources, though the reduced resources help with routability and performance for high values of \(w\). FLiMS wins in both performance and resource utilisation by a great margin, while FLiMSj lays between FLiMS and the alternatives WMS/EHMS in most aspects to offer the additional functionality of section \ref{flimsj}.

This evaluation focused more on the merging techniques, ignoring the building of merge trees or the handling of the list endings. The relevant paper \cite{fsorter} elaborates on a complete sorter implementation based on FLiMS, with a highly competitive logic and time complexity combination over related work on complete sorting. %

\section{Software implementation using SIMD intrinsics}\label{avxflims}

The goal of this section is to experimentally show that a single-instruction multiple data (SIMD)-accelerated merge sort function based on FLiMS can compete with alternative popular sorting functions based on different sorting algorithms. Today's general purpose processors (CPUs) feature vector or SIMD instructions as a way to increase performance in numerous compute and memory-intensive applications. Parallel merging algorithms implemented using SIMD-intrinsics, have already been shown to improve the sorting performance on CPUs. 

Chhugani et al. \cite{simd2008} used the rather simple merge algorithm based on the bitonic merger (see figure \ref{mergefirstl}) to enable high-throughput merging on an older Intel processor. %
Since it uses a full (non-partial) bitonic merger, both the lower \(w\) and upper \(w\) are used. %
As the upper \(w\) is calculated after \(log_2(2w)\) stages, this could have the feedback problem that is addressed by the latest research on FPGA merging (as summarised in table \ref{flimscompt}). However, on CPUs this is not much of a concern, because the pipelining functionality is not that advanced to achieve efficient task-pipelining. For example, for a single layer of compare-and-swap (\emph{CAS}) units, there need to be at least three SIMD intrinsics, one for \(min\), one for \(max\) and at least one \emph{shuffle}. The latter is to appropriately permute the inputs to emulate the \emph{CAS} network topology. %

In practice, how FLiMS could help is with a reduction in instruction count, as the lower \(w\) of the result is not needed. In other words, with roughly a similar number of instructions, FLiMS can merge with double the amount of parallelism \(w\). Other desirable characteristics of FLiMS on SIMD are the bitonic merge topology, which is more regular/ symmetric when compared with the odd-even merge sort topology, and the elimination of the rotation of the inputs. A similar discussion can be made for comparing with the other merger alternatives of section \ref{flimscomp}, which require more comparisons and lengthier pipelines than FLiMS. %

In order to assess the efficiency of FLiMS as an SIMD algorithm, a manually vectorised code in C++ %
is developed using Advanced Vector Extensions for 256-bit registers (AVX2). %
It is then extended to perform full sorting, %
and is finally %
compared to other existing sorting functions.%

\subsection{Merge function implementation}\label{flimssae}%

An SIMD-based implementation of FLiMS can be split into two main parts; one for the \emph{MAX} (selector) stage and one for the butterfly network. The \emph{MAX} stage is responsible for fetching the next elements from the input lists, selecting the top \(w\) and feeding it to the butterfly network. The butterfly network is more straightforward to implement using intrinsics, as it is a part of the bitonic sorter, which was already explored in a similar context \cite{bramas}. 

The \emph{MAX} stage can be implemented in SIMD by keeping the \(cA\) and \(cB\) in vector registers and generating the top \(w\) by comparing them. There are two ways to implement its fetching functionality. The first is to keep and update \(w\) independent pointers per input representing the input queues. Then mask/gather intrinsics are called to update the \(cA\) and \(cB\) values which are kept in vector registers. This is more faithful to the original FLiMS algorithm, but it is less efficient to use a gather AVX2 intrinsic \cite{intel} for fetching otherwise continuous memory locations. The preferred faster method involves ``pre''-fetching \(w\)-sized batches of elements, which is reminiscent of FLiMSj of section \ref{flimsj}. %

One complication with fetching whole vectors for the \emph{MAX} stage is that the \(cB\) vector needs to be %
in reverse order. This is done by the intrinsic \emph{\_mm256\_permutevar8x32\_epi32()} to reverse the contents of each batch fetched from list B. %
The comparison of the \emph{MAX} units is done by the intrinsic \emph{\_mm256\_cmpgt\_epi32()} and the result boolean vector is also used to fetch the next heads selectively after blending \emph{\_mm256\_blendv\_epi8()} with the current set of heads. The result of the compare instruction  \emph{\_mm256\_cmpgt\_epi32()} has the form of zeros and ones, each now denoting the source list of the next heads. The negation \emph{\_mm256\_andnot\_si256()} of this vector is also used to get the equivalent vector to be used when blending the next vector from the second input. %

With respect to the butterfly network part, %
each layer of \emph{CAS} units can be calculated by calling both a min (\emph{\_mm256\_min\_epi32()})  and a max (\emph{\_mm256\_max\_epi32()}) instruction consecutively, since each \emph{CAS} has two outputs. Alternatively, a compare instruction can be used once \emph{\_mm256\_cmpgt\_epi32()}, but there is a performance overhead from the additional calls of the \emph{\_mm256\_blendv\_epi8()} instruction, required to translate the result vector of \emph{\_mm256\_cmpgt\_epi32()} into minimums and maximums. Before each \emph{CAS} layer, the inputs must be properly aligned, and this is done through different permute intrinsics such as \emph{\_mm256\_shuffle\_epi32()} and \emph{\_mm256\_permute2x128\_si256()}.

In principle, emulating a wider FLiMS, can result in increased data locality, as more computation is done in registers. On the other hand, more logic needs to be represented, and the instruction count of the loop increases, with the additional disadvantage of less-obvious data dependencies. Since the number of vector units are limited in processors, emulating more logic on the merge loop can trigger more cache accesses and worsen the performance. 

\begin{figure}[h!]
\centering
\includegraphics[scale=0.85, trim=0 10 0 10 ]{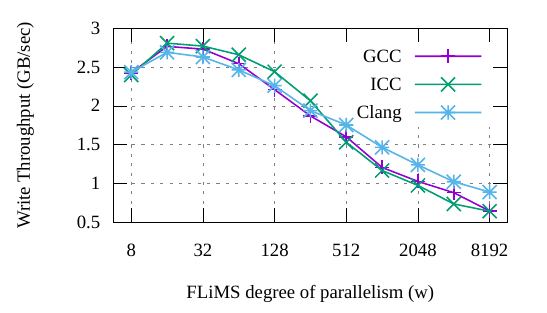}
\caption{Finding the optimal \(w\) value for AVX2-based FLiMS} \label{simdsortf3}
\end{figure}

Figure \ref{simdsortf3} shows how the degree of parallelism \(w\) of the emulated FLiMS influences the achievable throughput on an AVX2-native processor (Intel i7-8809G, with a steady 4.2 GHz clock). The C++ code for the 2-way merge function is generated by a python script, and is compiled with \(-O3\) and \(-march=native\). Two sorted random inputs of \(2^{24}\) elements are fed into the FLiMS merge function. The  conclusion is that at \(w=16\) and \(32\) the throughput is the highest, and there is little variation between different compilers.

\subsection{Complete sorting %
}\label{csfev}

The FLiMS-based CPU merge function can be used recursively to merge input of arbitrary length, by accelerating merge sort with SIMD instructions. 

As a complement to the FLiMS-based merge function, a sort-in-chunks function is developed to facilitate the need for initial sorted chunks, as well as to provide long-enough chunks for FLiMS to benefit from streaming access patterns.
This function is based on the bitonic sorter, with a similar technique to section \ref{flimssae} for building the butterfly network. A similar approach was followed in an SIMD-based quicksort implementation \cite{bramas}, although here it is implemented from scratch with %and optimised for our use case with %
AVX2 intrinsics. %
In our use case, the optimal sorted chunk size is found to be 512 integers. % for the least runtime.

The performance of the FLiMS-based sort function is compared against the C++ Standard Library implementation of sorting \emph{std::sort()}, as well as a highly-optimised \cite{bramas} SIMD-based radix sort implementation from Intel’s integrated performance primitives (IPP).

Additionally, a multi-threaded version of the FLiMS-based SIMD sort function is implemented, with the help of OpenMP pragmas. Both the merging function and the sort-in-chunks function are unaltered. The parallelisation is done on the calls of each, operating on equally-sized consecutive portions of the entire input, when possible. %The sorting-in-chunks now happens on all cores, operating on equally-sized consecutive portions of the entire input. A similar loop initiates multiple instances of the FLiMS-based merge, as long as there are enough sublists in the current merge iteration. %

The performance of the multi-threaded FLiMS-based sort is compared against the single-threaded baselines, as well as a parallel sort implementation in the Boost C++ libraries. The \emph{block\_indirect\_sort()} function implements the samplesort sorting algorithm, and is regarded as one of the best performing C++ sort implementations \cite{boost}.

\begin{figure}[h!]
\centering
\includegraphics[scale=0.85, trim=7 7 -7 10 ]{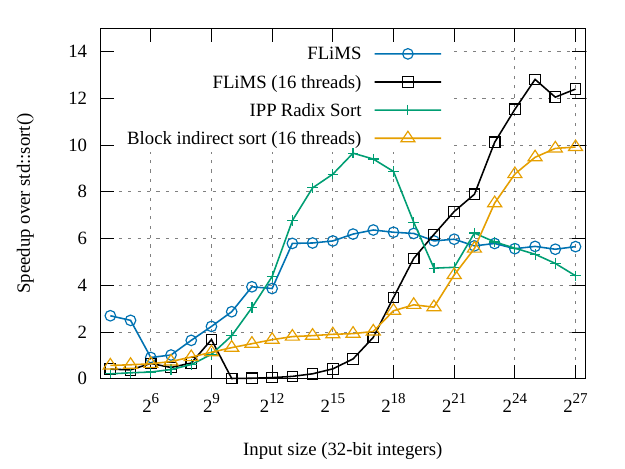}
\caption{Evaluating the FLiMS-based SIMD complete sort C++ function}\label{simdsortf2}
\end{figure}

Figure \ref{simdsortf2} presents the results of both the single-threaded and multi-threaded experiments. The target processor is the 16-thread Ryzen Pro 4750U. %
The main observation is that the 16-thread FLiMS sort function surpasses the performance of the 16-thread \emph{block\_indirect\_sort()} for the input range from \(2^{17}\) to \(2^{27}\). A hybrid approach can be used to enable the single-threaded version of FLiMS for below \(2^{20}\) to achieve the highest-performance overall, except radix sort. 

Intel's radix sort takes the lead between \(2^{12}\) and \(2^{19}\) on this AMD processor, but IPP radix sort has some notable limitations. These include the less predictable performance across different systems when run across different processors (not demonstrated, for brevity), a restrictive license and radix-sort related implications, such as a limit of the input length to about \(2^{28}\) for the tested implementation. %

As a conclusion, FLiMS-based sorting %
is an attractive approach to accelerate merge sort using SIMD intrinsics. However, %
it is still relevant to research solutions based on a variety of sorting methods,
as different algorithms are more appropriate in different distributions and use cases \cite{auger2015merge}. For instance, 
radix sort can perform fewer data passes on data of a restricted range, such as for 10-bit integers. %

\section{Future work}
Current and future work includes developing efficient FLiMS implementations targeting other technologies such as GPUs, and exploring the adoption of FLiMS in various applications such as database analytics \cite{fsorter, mergejoin} and beyond. 

One concern in today's FPGA research on sorting is that it is mostly limited to fixed-width values \cite{thiemj}. Therefore, it would be helpful to also study FLiMS adaptations or alternatives for data of arbitrary width, such as strings \cite{asiatici2021many}.

The SIMD evaluation could also be extended with additional optimisation for specific processors and data. This could also include the skewness and stable variations and their applicability in this context. An AVX-512 version has already been developed, but for the target processor (Xeon 8124M) the performance benefits over AVX2 on the same processor were underwhelming. This possibly related to the efficiency of AVX-512 in the specific micro-architecture, though it would be appropriate to further investigate the applicability of FLiMS in future processor technologies. %

It would also be interesting to formally prove any optimalities FLiMS may exhibit, as well as to try to find equivalent circuits for merging more than two sorted input lists, for further reducing the size of parallel merge trees.

\section{Conclusions}

FLiMS is currently the most hardware-efficient 2-way merge block on FPGAs. It can be used to build efficient high-throughput merge trees for facilitating sorting of unsorted input on hardware. It features fewer and/or simpler pipeline stages than the alternatives, while achieving a higher amount of parallelism with less hardware resource utilisation. The skewness optimisation is the equivalent workaround found in an older merger, while maintaining the decentralised nature of the \emph{MAX} entities logic, which is collectively used as a scalable selector logic. A variation for implementing stable sort is also presented to facilitate the needs of some database applications, as well as FLiMSj for unifying the dequeue signals that can be costly in some memory configurations. An SIMD implementation of FLiMS and its multi-threaded variant on a modern processor are also found to outperform popular highly-optimised C++ sort  libraries. %
Sorting using such a high-throughput merge block can be more appropriate for big data applications than alternative approaches, %
 since it yields streaming memory access patterns and can also be applied recursively for arbitrarily long data without keeping growing states.

\section*{Acknowledgments}
\footnotesize
This research was sponsored by dunnhumby. The support of Microsoft and the United Kingdom EPSRC (grant number EP/L016796/1, EP/I012036/1, EP/L00058X/1, EP/N031768/1 and EP/K034448/1), European Union Horizon 2020 Research and Innovation Programme (grant number 671653) is gratefully acknowledged. 

\ifCLASSOPTIONcaptionsoff
  \newpage
\fi
\small

\AtNextBibliography{\footnotesize}
\printbibliography

@misc{intel,
  title={Intel Intrinsics Guide},
  author={{Intel (R)}},
  url={https://software.intel.com/sites/landingpage/IntrinsicsGuide/}
}

@inproceedings{ehms,
  title={Towards an Efficient Hardware Architecture for Odd-Even Based Merge Sorter},
  author={Elsayed, Elsayed A and Kise, Kenji},
  booktitle={2019 IEEE 13th Int’l Symp. on Embedded Multicore/Many-core Systems-on-Chip (MCSoC)},
  pages={249--256},
  year={2019},
  organization={IEEE}
}

@inproceedings{vms,
  title={Very Massive Hardware Merge Sorter},
  author={Saitoh, Makoto and Kise, Kenji},
  booktitle={2018 Int’l Conference on Field-Programmable Technology (FPT)},
  pages={86--93},
  year={2018},
  organization={IEEE}
}

@inproceedings{mergejoin,
  title={Accelerating the merge phase of sort-merge join},
  author={Papaphilippou, Philippos and Pirk, Holger and Luk, Wayne},
  booktitle={29th Int’l Conference on Field Programmable Logic and Applications (FPL)},
  pages={100--105},
  year={2019},
  organization={IEEE}
}

@inproceedings{casper,
  title={Hardware acceleration of database operations},
  author={Casper, Jared and Olukotun, Kunle},
  booktitle={the 2014 ACM/SIGDA Int’l Symp. on Field--programmable gate arrays},
  pages={151--160},
  year={2014}  
}

@inproceedings{batcher,
  title={Sorting networks and their applications},
  author={Batcher, Kenneth E},
  booktitle={the April 30--May 2, 1968, spring joint computer conference},
  year={1968},
  organization={ACM}
}

@inproceedings{flims,
  title={{FLiMS: Fast Lightweight Merge Sorter}},
  author={Papaphilippou, Philippos and Brooks, Chris and Luk, Wayne},
  booktitle={2018 Int’l Conference on Field-Programmable Technology (FPT)},
  pages={78--85},
  year={2018},
  organization={IEEE}
}

@inproceedings{mms,
  title={A High-Performance and Cost-Effective Hardware Merge Sorter without Feedback Datapath},
  author={Saitoh, Makoto and Elsayed, Elsayed A and Van Chu, Thiem and Mashimo, Susumu and Kise, Kenji},
  booktitle={2018 IEEE 26th Annual Int’l Symp. on Field-Programmable Custom Computing Machines (FCCM)},
  pages={197--204},
  year={2018},
  organization={IEEE}
}

@inproceedings{pmt,
  title={Parallel hardware merge sorter},
  author={Song, Wei and Koch, Dirk and Luj{\'a}n, Mikel and Garside, Jim},
  booktitle={24th Annual Int’l Symp. on Field-Programmable Custom Computing Machines (FCCM)},
  pages={95--102},
  year={2016},
  organization={IEEE}
}

@inproceedings{mashimo2017high,
  title={High-Performance Hardware Merge Sorter},
  author={Mashimo, Susumu and Van Chu, Thiem and Kise, Kenji},
  booktitle={25th Annual Int’l Symp. on Field-Programmable Custom Computing Machines (FCCM)},
  year={2017},
  organization={IEEE}
}

@article{zachmann2013adaptive,
  title={Adaptive bitonic sorting},
  author={Zachmann, Gabriel},
  journal={Encyclopedia of Parallel Computing, David Padua, Ed},
  pages={146--157},
  year={2013}
}

@article{simd2008,
  title={{Efficient implementation of sorting on multi-core SIMD CPU architecture}},
  author={Chhugani, Jatin and Nguyen, Anthony D and Lee, Victor W and Macy, William and Hagog, Mostafa and Chen, Yen-Kuang and Baransi, Akram and Kumar, Sanjeev and Dubey, Pradeep},
  journal={the VLDB Endowment},
  volume={1},
  number={2},
  pages={1313--1324},
  year={2008},
  publisher={VLDB Endowment}
}

@conference{christopher,
title = {{Large Utility Sorting on FPGAs}},
author = "Kristiyan Manev and Dirk Koch",
year = "2018",
booktitle={Int’l Conf. on Field-Programmable Technology (FPT)},
organization={IEEE}
}

@inproceedings{fpgasort,
  title={{FPGASort: A high performance sorting architecture exploiting run-time reconfiguration on FPGAs for large problem sorting}},
  author={Koch, Dirk and Torresen, Jim},
  booktitle={the 19th ACM/SIGDA Int’l Symp. on Field programmable gate arrays},
  pages={45--54},
  year={2011}
}

@inproceedings{leaf,
  title={{A cost-effective and scalable merge sorter tree on FPGAs}},
  author={Usui, Takuma and Van Chu, Thiem and Kise, Kenji},
  booktitle={2016 Fourth Int’l Symp. on Computing and Networking (CANDAR)},
  pages={47--56},
  year={2016},
  organization={IEEE}
}

@article{linear,
  title={A shift register architecture for high-speed data sorting},
  author={Lee, Chen-Yi and Tsai, Jer-Min},
  journal={Journal of VLSI signal processing systems for signal, image and video technology},
  volume={11},
  number={3},
  pages={273--280},
  year={1995},
  publisher={Springer}
}

@article{bramas,
  title={{A novel hybrid quicksort algorithm vectorized using AVX-512 on Intel Skylake}},
  author={Bramas, Berenger},
  journal={Int’l Journal of Advanced Computer Science and Applications (IJACSA)},
  volume={8},
  number={10},
  pages={337--344},
  year={2017},
  publisher={SAI Society}
}

@inproceedings{kobayashi2015face,
  title={Face: Fast and customizable sorting accelerator for heterogeneous many-core systems},
  author={Kobayashi, Ryohei and Kise, Kenji},
  booktitle={2015 IEEE 9th Int’l Symp. on Embedded Multicore/Many-core Systems-on-Chip},
  pages={49--56},
  year={2015},
  organization={IEEE}
}

@inproceedings{pswitch,
author = {Papaphilippou, Philippos and Meng, Jiuxi and Luk, Wayne},
title = {{High-Performance FPGA Network Switch Architecture}},
year = {2020},
publisher = {ACM},
booktitle = {Int’l Symp. on Field-Programmable Gate Arrays (FPGA)},
pages = {76--85},
numpages = {10},
}

@inproceedings{fsorter,
  title={{An Adaptable High-Throughput FPGA Merge Sorter for Accelerating Database Analytics}},
  author={Papaphilippou, Philippos and Brooks, Chris and Luk, Wayne},
  booktitle={2020 30th Int’l Conference on Field Programmable Logic and Applications (FPL)},
  pages={65--72},
  year={2020},
  organization={IEEE}
}

@article{farmahini2012modular,
  title={Modular design of high-throughput, low-latency sorting units},
  author={Farmahini-Farahani, Amin and Duwe III, Henry J and Schulte, Michael J and Compton, Katherine},
  journal={IEEE Transactions on Computers},
  volume={62},
  number={7},
  pages={1389--1402},
  year={2012},
  publisher={IEEE}
}

@misc{boost,	
	title = {Boost.Sort 3.- Parallel Algorithms},
	booktitle = {Boost Develop Library Documentation},
	url={https://www.boost.org/doc/libs/develop/libs/sort/doc/html/sort/parallel.html},
	note = "[Online; accessed 9-April-2021]"
}

@article{auger2015merge,
  title={Merge strategies: from merge sort to Timsort},
  author={Auger, Nicolas and Nicaud, Cyril and Pivoteau, Carine},
  url = {https://hal-upec-upem.archives-ouvertes.fr/hal-01212839},
  journal = {working paper or preprint},
  year={2015}
}

@article{ehmsp,
  title={High-Performance and Hardware-Efficient Odd-Even Based Merge Sorter},
  author={Elsayed, Elsayed A and Kise, Kenji},
  journal={IEICE Transactions on Information and Systems},
  volume={103},
  number={12},
  pages={2504--2517},
  year={2020},
  publisher={The Institute of Electronics, Information and Communication Engineers}
}

@misc{thiemj,  
  title = {\jap{FPGA を使って基本的なアルゴリズムのソーティングを劇的に高速化 (5)}},
  author = {Van Chu, Thiem},
  howpublished = "\url{https://www.acri.c.titech.ac.jp/wordpress/archives/4713}",
  year = {2020}, 
  note = "[Online; accessed 10-November-2021]"
}

@article{asiatici2021many,
  title={How many CPU cores is an FPGA worth? Lessons learned from accelerating string sorting on a CPU-FPGA system},
  author={Asiatici, Mikhail and Maiorano, Damian and Ienne, Paolo},
  journal={Journal of Signal Processing Systems},
  pages={1--13},
  year={2021},
  publisher={Springer}
}

@misc{cullen,  
  title = {{Cullen numbers: n * 2\^\ n + 1.}},
  author = {{The On-Line Encyclopedia of Integer Sequences}},
  howpublished = "\url{https://oeis.org/A002064}",
  note = "[Online; accessed 3-June-2021]"
}
\vspace{-8em}

\begin{IEEEbiography}[{\includegraphics[width=1in, height=1.25in,clip, keepaspectratio]{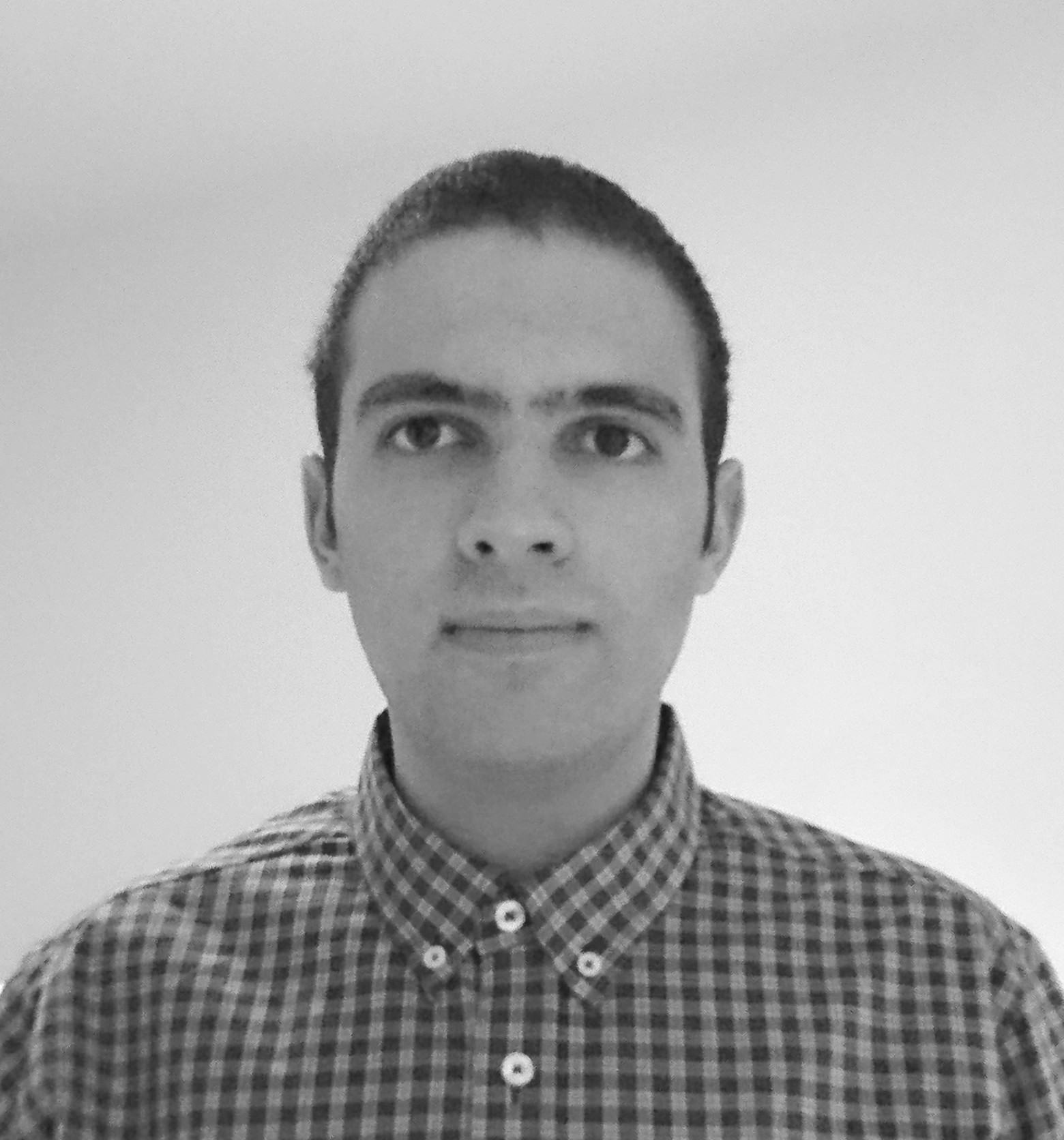}}]{Philippos Papaphilippou}
got his PhD from Imperial College London. His PhD was
funded by dunnhumby for researching novel accelerators to improve the
performance of big data analytics. His research topics include FPGAs, sorting algorithms,
network switches, multi-processor architecture and data science.
\vspace{-9em}
\end{IEEEbiography}

\begin{IEEEbiography}[{\includegraphics[width=1in,height=1.25in,clip,keepaspectratio]{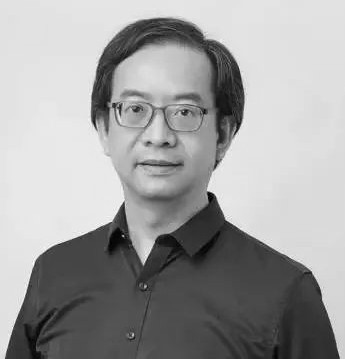}}]{Wayne Luk}
is a professor of computer engineering at Imperial College London. He leads
the Programming Languages and Systems Section, and the Custom Computing Research
Group at the Department of Computing. He is a Fellow of the Royal Academy of
Engineering, the IEEE, and the BCS.
\vspace{-9em}
\end{IEEEbiography}

\begin{IEEEbiography}[{\includegraphics[width=1in,height=1.25in,clip,keepaspectratio]{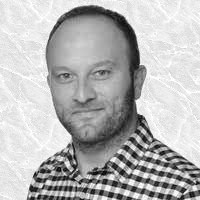}}]{Chris Brooks}
is the Head of Science Innovation at dunnhumby, UK.  He is accountable for researching and developing new science techniques and technical implementations.  He leads science research into a variety of areas, focussed primarily on the retail domain.
\vspace{-9em}
\end{IEEEbiography}

\end{document}